\documentstyle[epsf,twoside,fleqn,espcrc2]{article}

\newcommand{\AmS}{{\protect\the\textfont2
  A\kern-.1667em\lower.5ex\hbox{M}\kern-.125emS}}

\def\simm#1{\mathop{\vtop{\ialign{##\crcr
        $\hfil\displaystyle{#1}\hfil$\crcr\noalign{\kern0.5pt\nointerlineskip}
        $\sim$\crcr\noalign{\kern0.5pt}}}}\limits}

\def\be{\begin{equation}}
\def\ee{\end{equation}}
\def\ba{\begin{eqnarray}}
\def\ea{\end{eqnarray}}

\def\P{{\sf \bf P}}
\def\R{{\sf \bf R}}
\def\W{{\sf \bf W}}
\def\C{{\sf \bf C}}
\def\PW{{\sf \bf PW}}
\def\PC{{\sf \bf PC}}
\def\RW{{\sf \bf RW}}
\def\RC{{\sf \bf RC}}

\title{ 
\vspace*{-35pt}
{\normalsize \hfill {\sf UTCCP-P-28}} \\
\vspace*{-6pt}
{\normalsize \hfill {\sf Oct.\ 1997}} \\
Hadron spectroscopy and static quark potential in full QCD: \\
A comparison of improved actions on the CP-PACS%
\thanks{ Talks presented by R.~Burkhalter and T.~Kaneko at \mbox{Lattice97},
Edinburgh, Scotland, 22--26 July 1997.}  
}

\author{
CP-PACS Collaboration:\\
S.\ Aoki\rlap,\address{Institute of Physics,
University of Tsukuba, Tsukuba, Ibaraki 305, Japan}
G.\ Boyd\rlap,\address{Center for Computational Physics, 
University of Tsukuba, Tsukuba, Ibaraki 305, Japan}
R.\ Burkhalter\rlap,$^{\rm b}$
S.\ Hashimoto\rlap,\address{Computing Research Center, 
High Energy Accelerator Research Organization (KEK), Tsukuba, Ibaraki 305, 
Japan}
N.\ Ishizuka\rlap,$^{\rm a,b}$
Y.\ Iwasaki\rlap,$^{\rm a,b}$
K.\ Kanaya\rlap,$^{\rm a,b}$
T.\ Kaneko\rlap,$^{\rm a}$
Y.\ Kuramashi\rlap,\address{Institute of Particle and Nuclear Studies,
High Energy Accelerator Research Organization (KEK), Tsukuba, Ibaraki 305,
Japan}
M.\ Okawa\rlap,$^{\rm d}$
A.\ Ukawa\rlap,$^{\rm a}$
and
T.\ Yoshi\'e$^{\rm a,b}$
}

\begin{document}
\renewcommand{\textfraction}{0.1}
\renewcommand{\topfraction}{0.9}

\begin{abstract}

We present first results from a full QCD calculation on the CP-PACS,
comparing various actions at $a^{-1} \sim 1 {\rm GeV}$ and $m_\pi/m_\rho
\approx 0.7$--0.9. We use the plaquette and a
renormalization group improved action for the gluons, and 
the Wilson and the SW-Clover action for quarks.  We find that significant
improvements in the hadron spectrum results from improving the quarks,
while the gluon improvement is required for a rotationally invariant static 
potential. An ongoing effort towards exploring the chiral limit in full 
QCD is described.

\end{abstract}

\maketitle

\section{Introduction}

With the progress in recent years of quenched simulations of QCD,
deviations of the quenched hadron spectrum from experiment are becoming 
apparent. 
Most recently the precision quenched results from the CP-PACS show 
a systematic deviation not only in the meson sector, but also in 
the spectrum of baryons\cite{KanayaLat97}.  Clearly the time
has come to bolster efforts towards simulations of full QCD.

Full QCD simulations are, however, extremely computer time consuming
compared to those of quenched QCD.  This leads us to consider improved
actions for our simulation of full QCD.

Studies of improved actions have been widely pursued in the last few years. 
However, a systematic investigation of how various terms added to the gauge
and quark actions, taken separately, affect light hadron observables has
not been made in full QCD.  In this article we report results of
such a study carried out on the CP-PACS.

\section{Actions and simulation parameters}

\begin{table}[t]
\setlength{\tabcolsep}{0.2pc}
\caption{Simulation parameters. All the runs are on a $12^3\!\times\!32$
lattice, except (*), which is on a $16^3\!\times\!32$ lattice. For runs
marked with ($\dag$) the static quark potential is measured. Errors in the
mass ratios are statistical but for $a^{-1}$ they are systematic, taking into
account uncertainties from the chiral extra\-polation.}
\label{tab:param}
\begin{tabular}{clllrll}
\hline
 & $\beta$ & $K$ & $c_{SW}$ & \# $\:$ & $m_\pi/m_\rho$ & $a^{-1}$\\
 &         &     &          & conf &                & GeV     \\
\hline
\PW  & 4.8        & .1846 &        & 222    & .829(1)  & \\ 
     &            & .1874 &        & 200    & .772(2)  & \\
     &            & .1891 &        & 200    & .703(3)  & 0.99(1)\\
     \cline{2-7}
     & $5.0^{\dag}$ & .1779 &        & 300    & .846(1)  & \\
     &              & .1798 &        & 301    & .794(2)  & \\
     &              & .1811 &        & 301    & .715(4)  & 1.13(5)\\
\hline
\RW  & 1.9        & .1632 &        & 200    & .899(1)  & \\
     &            & $.1688^{\dag}$ &        & 200    & .803(3)  &\\
     &            & .1713 &        & 200    & .689(4)  & 1.22(8)\\
     \cline{2-7}
     & $2.0^{\dag}$ & .1583 &        & 300    & .898(1)  & \\
     &              & .1623 &        & 300    & .828(2)  & \\
     &              & .1644 &        & 305    & .739(3)  & 1.36(8)\\
                    \cline{3-7}
\hline
\end{tabular}
\vspace{-5mm}
\end{table}
\setcounter{table}{0}

\begin{table}[t]
\setlength{\tabcolsep}{0.2pc}
\caption{Continued.}
\begin{tabular}{clllrll}
\hline
\PC  & 5.0        & .1590 & 1.0    & 100    & .826(2)   & \\
     &            & .1610 & 1.0    & 100    & .791(4)   & \\
     &            & .1630 & 1.0    & 101    & .712(6)   & 0.91(2)\\
                  \cline{3-7}
     & $5.0^{\dag}$ & .1415 & 1.855  & 200    & .810(3)   & \\
     &              & .1441 & 1.825  & 200    & .757(4)   & \\
     &              & .1455 & 1.805  & 200    & .714(4)   & 0.88(6)\\
     \cline{2-7}
     & $5.2^{\dag}$ & .1390 & 1.69   & 248    & .840(2)  & \\ 
     &              & .1410 & 1.655  & 232    & .794(3)  & \\
     &              & .1420 & 1.64   & 200    & .724(8)  & 1.40(15) \\
     \cline{2-7}
     & 5.25       & .1390 & 1.637  & 198    & .835(3)  & \\
     &            & .1410 & 1.61   & 194    & .758(6)  & 1.5(1)\\
\hline
\RC  & $1.9^*$    & .1370 & 1.55   & 203    & .845(1) & \\  
     &            & .1400 & 1.55   & 198    & .779(2) & \\
     &            & $.1420^{\dag}$ & 1.55   & 202  & .690(3) & \\
     &            & $.1430^{\dag}$ & 1.55   & 212  & .611(4)   \\
     &            & .1435 & 1.55   & 263    & .544(5) &  \\
     &            & .1440 & 1.55   &  79    & .41(1) & 1.19(6) \\
     \cline{2-7}
     & $1.9^{\dag}$ & .1370 & 1.55   & 267    & .846(2)  & \\ 
     &              & .1400 & 1.55   & 214    & .776(2)  & \\
     &              & .1420 & 1.55   & 268    & .684(3)  & 0.97(10)\\
     \cline{2-7}
     & 2.0        & .1420 & 1.0    & 100    & .878(1)  & \\
     &            & .1450 & 1.0    & 100    & .830(2)  & \\
     &            & .1480 & 1.0    & 100    & .710(6)  & 1.24(9)\\
                  \cline{3-7}
     & 2.0        & .1300 & 1.505  & 100    & .910(1)  & \\  
     &            & .1370 & 1.505  &  90    & .794(3)  & \\
     &            & .1388 & 1.505  &  90    & .710(8)  & 1.35(20)\\
                  \cline{3-7}
     & $2.0^{\dag}$ & .1300 & 1.54   & 201    & .902(1)  & \\
     &              & .1340 & 1.529  & 200    & .862(2)  & \\
     &              & .1370 & 1.52   & 200    & .791(3)  & \\
     &              & .1388 & 1.515  & 200    & .700(6)  & 1.35(15)\\
\hline
\end{tabular}
\vspace{-5mm}
\end{table}

We are carrying out a comparative study of the light hadron spectrum and static
quark potential for four action combinations. For gluons we choose either
the standard plaquette action (\P) or an action (\R) which was obtained by
a renormalization group treatment \cite{Iwasaki83} and includes a
rectangular Wilson loop. For the quarks we use either the Wilson action
(\W) or the Clover action (\C)~\cite{clover}. For the clover coefficient
$c_{SW}$ we compare:\\ 
\noindent 
(a) the tree value $c_{SW}=1$.\\
\noindent 
(b) the meanfield (MF) improved value \cite{meanfield} $c_{SW} = P^{-3/4}$
with $P$ the self-consistently determined plaquette average, and hence 
a different value of $c_{SW}$ for each $K$.\\
\noindent 
(c) a perturbative meanfield (pMF) improved value 
$c_{SW}=(1-0.8412\cdot\beta^{-1})^{-3/4}$, employed for the \R\ action, 
where the one-loop value is used for the plaquette.\\
We expect the extent of improvement to be clearer at a coarser lattice
spacing.  We therefore attempt to tune the coupling constant $\beta$ so
that the lattice spacing equals $a^{-1} \sim 1$GeV.

Our simulations are carried out for two flavors of quarks, mostly on a
$12^3\times 32$ lattice. We employ the hybrid Monte Carlo algorithm to
generate full QCD configurations, generally at three values of $K$
corresponding to $m_\pi/m_\rho \approx 0.7$-0.9. Details of the simulation
parameters are given in Table~\ref{tab:param}. For the inversion of the
fermion matrix we first used the MR algorithm but switched later to
BiCGStab. The molecular dynamics step size is chosen to yield an
acceptance of 70-90\%.

We measure hadron propagators every 5 trajectories, using smeared sources
and point sinks following the method of our quenched study
\cite{KanayaLat97}. Errors of masses are determined with a jackknife 
analysis using a bin size of 5. The lattice spacing is fixed by 
the $\rho$ meson mass extrapolated to the chiral limit through the 
form $m_\rho=m_\rho^{(0)}+c_2m_\pi^2$ or $m_\rho^{(0)}+c_2m_\pi^2+c_3m_\pi^3$.
The errors for $a^{-1}$ quoted in Table~\ref{tab:param} include the 
systematic errors in the extrapolation, estimated from the two fitting functions
above, and by varying the fitting ranges.

Measurements of the static quark potential are performed on about $100$
configurations at those parameters marked with a $\dag$ in
Table~\ref{tab:param}. In these measurements we employ the smearing
procedure proposed in Ref.~\cite{PG-StrTns} and extract the ground state
potential in the same way as in Ref.~\cite{PG-Tcs}.

\section{Light hadron spectrum}
\label{sec:spectrum}
\begin{figure}[htb]
\vspace{-4mm}
\begin{center}
\leavevmode
\epsfxsize=8cm
\epsfbox{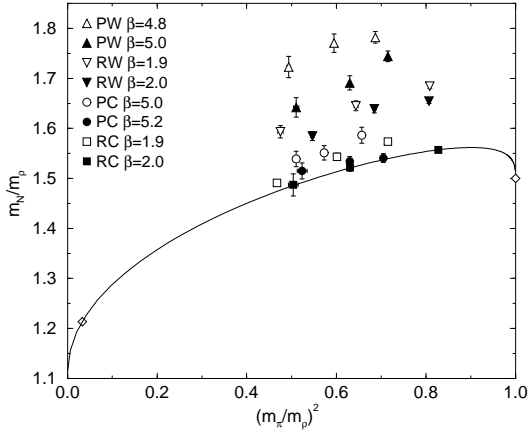}
\end{center}
\vspace{-15mm}
\caption{$m_N/m_\rho$ as a function of $(m_\pi/m_\rho)^2$ 
for four combinations of the action. Clover results are for the 
choice $c_{SW}={\rm MF}$.}
\label{fig:EPN}
\vspace{-6mm}
\end{figure}
\begin{figure}[htb]
\vspace{-5mm}
\begin{center}
\leavevmode
\epsfxsize=8cm
\epsfbox{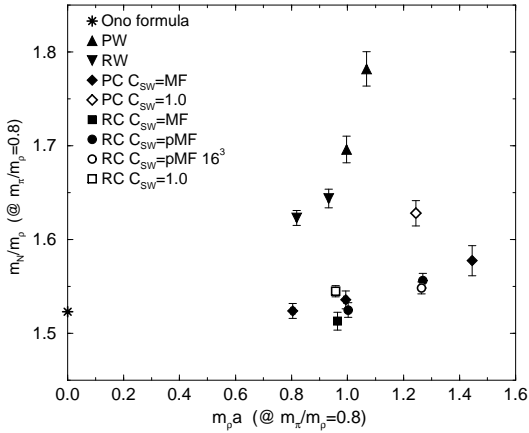}
\end{center}
\vspace{-15mm}
\caption{Scaling
behaviour of $m_N/m_\rho$ at fixed $m_\pi/m_\rho$ as a function of $m_\rho a$
for various combinations of the action. The star on the left is the prediction 
of the phenomenological formula.}
\label{fig:scaling}
\vspace{-6mm}
\end{figure}

\begin{figure}[t]
\vspace{-7mm}
\begin{center}
\leavevmode
\epsfxsize=8cm
\epsfbox{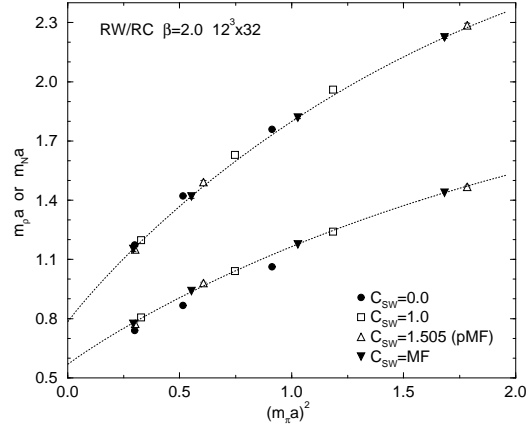}
\end{center}
\vspace{-15mm}
\caption{ $m_\rho$ and $m_N$ as a function of $m_\pi^2$ for different
choices of $c_{SW}$. Dashed lines are chiral fits, cubic in $m_\pi$, 
to data with $c_{SW}={\rm MF}.$
}
\label{fig:chiral}
\vspace{-6mm}
\end{figure}

Our main results for the effect of improved actions on hadron masses are
displayed in Fig.~\ref{fig:EPN}, in which the ratio $m_N/m_\rho$ is plotted
as a function of $(m_\pi/m_\rho)^2$ for the four action combinations.  The
solid curve represents the well-known phenomenological mass formula
\cite{ONO}.  

For the standard action combination \PW\ (upward triangles), 
the ratios are well above the
phenomenological curve.  When we improve the gauge action (the \RW\ case, 
shown by downward triangles),
the data points come closer to the curve.  By far the most conspicuous
change, however, is observed when we introduce the clover term to the quark
action. For both the \PC\ (circles)  and \RC\ (squares) cases, 
the data points lie much nearer
to, or on top of, the phenomenological curve.

Looking only at Fig.~\ref{fig:EPN}, however, is misleading because the runs
for different action combinations are performed at slightly different
lattice spacings. In Fig.~\ref{fig:scaling} we therefore 
show the $a$-dependence. We interpolate the ratios $m_N/m_\rho$ to a
common, fixed ratio $m_\pi/m_\rho=0.8$ and plot  $m_N/m_\rho$ vs. $m_{\rho}$
in lattice units, with $m_{\rho}$ also interpolated to  $m_\pi/m_\rho=0.8$.
Now it becomes clear that the better results for the \RW\ case are
only due to a finer lattice spacing and that the \PW\ and \RW\ results 
seem to lie almost on the same curve. 
The results for the \PC\ and \RC\ cases with the
MF or pMF choice of $c_{SW}$ also lie on a common curve, which, however, is
much flatter. Somewhat less, but still quite substantial, improvement is
found for the tree level $c_{SW}$. The same trend is seen also at
$m_\pi/m_\rho=0.7$ and for the ratio $m_\Delta/m_\rho$
\cite{OurFullPaper}.

Changing the gluon and quark actions one at a time, we have been able to
see clearly that the clover term plays a decisive role in improving the
spectrum. In this regard, improving the gluon action has much less effect.

Another interesting feature in our hadron mass data is that they exhibit a
negative curvature in terms of $m_\pi^2$ towards the chiral limit. In
Fig.~\ref{fig:chiral} we show this effect for the $\rho$ meson and nucleon
masses using the \RW\ and \RC\ action combinations and several choices of
$c_{SW}$. We also note that for the Clover action the $\rho$ meson masses are
shifted to a higher value compared to those for the Wilson action. 
For the nucleon,
however, we see no difference between the Wilson and Clover cases. 
This confirms an
earlier observation using staggered dynamical fermions\cite{Collins}.

\section{Dispersion relation}
\label{sec:dispersion}
\begin{figure}[t]
\hspace*{-5mm}
\epsfxsize=8.5cm
\epsfbox{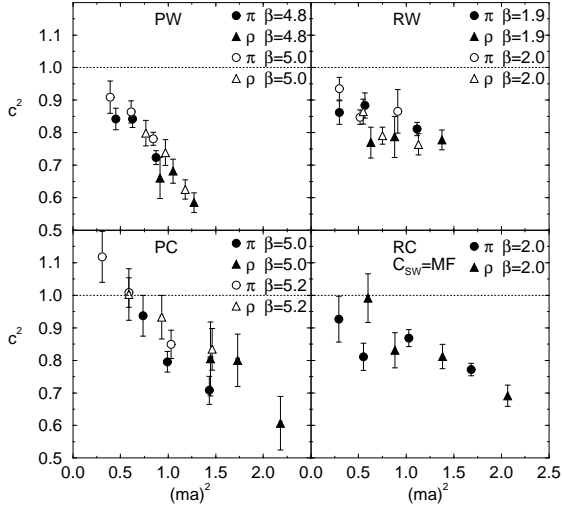}
\vspace{-10mm}
\caption{``Speed of light'' as a function of the meson mass for four action
combinations. Clover results are for the choice $c_{SW}={\rm MF}$}
\label{fig:c2}
\vspace{-6mm}
\end{figure}

Improving the action is expected to lead to a
dispersion relation closer to that in the continuum. As a
measure for the closeness to the continuum we define the ``speed of light''
$c$ through $c^2=dE(p)^2/dp^2|_{p=0}$.  Values of $c$ are extracted 
by a linear fit of $E(p)^2$ to $p^2$, using the
three lowest momenta.  
In the continuum $c$ equals one for all values of the rest mass. 
On the lattice, however, $c$ is only expected to go to one for
vanishing rest mass. For nonzero rest mass it is generally different from
one. An improved action is expected to have a ``speed of light'' closer to
one for nonzero rest masses. 

Our results for $c$ are plotted in Fig.~\ref{fig:c2}.  
An improvement is most clearly seen in a comparison of the \PW\ and \RC\
action combinations.  We are not able to tell 
whether the improvement 
is due more to the gauge or quark action, however, since 
the \RW\ and \PC\ cases both also show signs of similar improved behavior 
within the present statistics.


\section{Static quark potential}

\begin{figure}[t]
\hspace*{-5mm}
\epsfxsize=8.5cm
\epsfbox{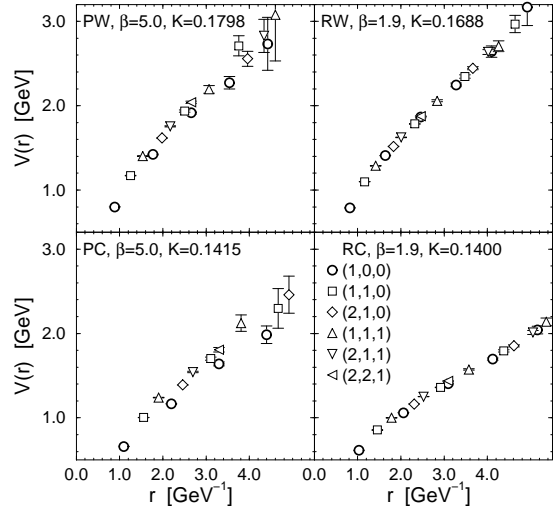}
\vspace{-10mm}
\caption{
Static quark potential for four action combinations at 
$m_{\pi}/m_{\rho} \simeq 0.8$. Scales are normalized
by the lattice spacing determined from $m_{\rho}$ in the chiral limit.
Different symbols correspond to potential data measured in different spatial 
directions along the vector indicated in the figure. Clover results 
are for the choice $c_{SW}={\rm MF}$}
\label{fig:Vcrv}
\vspace{-6mm}
\end{figure}

\subsection{Rotational symmetry}

The static quark potential provides a good measure of the effectiveness
of improving the action in restoring the
rotational symmetry.  In the quenched case it is known that 
the symmetry is well restored by improving the gauge 
action\cite{PG-Tcs,Cornell}.  
Our typical results for full QCD taken for a quark mass 
corresponding to $m_\pi/m_\rho \approx 0.8$ are plotted in Fig.~\ref{fig:Vcrv}
for each action combination.

We see quite clearly in Fig.~\ref{fig:Vcrv} that
the symmetry is remarkably well restored by improving only the gauge action
also in full QCD.  The effect of quark action improvement on the
restoration of the symmetry is not clear in our data.  

This may be explained by the fact that 
the clover improvement does not remove terms of the Wilson action breaking 
rotational symmetry, 
while the \R\ action for gluons has much reduced 
coefficients for the rotationally variant terms compared to the plaquette 
action.

\subsection{Matching of lattice scale}

\begin{figure}[t]
\hspace*{-5mm}
\epsfxsize=8.5cm
\epsfbox{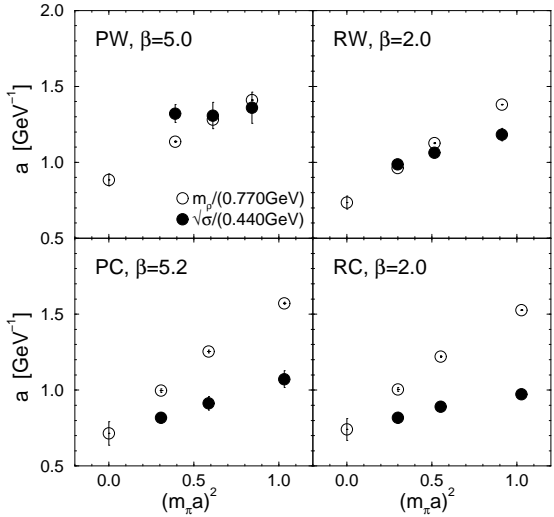}
\vspace{-10mm}
\caption{
Lattice spacing $a$ determined by the string tension $\sigma$ and the
$\rho$ meson mass as a function of $(m_{\pi}a)^2$ for the four 
action combinations. Clover data are for the choice $c_{SW}={\rm MF}$. 
Filled circles are obtained by identifying 
${\sigma}{a^2}=(440{\rm MeV})^2$, and open circles from  
${m_\rho}a=770$MeV at each values of $K$ and in the chiral limit.
}
\label{fig:a}
\vspace{-6mm}
\end{figure}

\begin{figure}[t]
\vspace{-4mm}
\begin{center}
    \leavevmode
    \epsfxsize=	8cm 
    \epsfbox{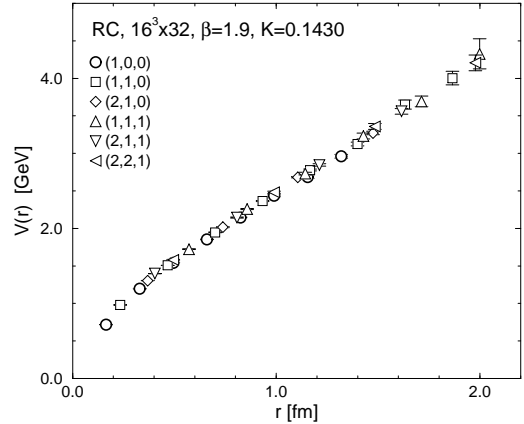}
\end{center}
\vspace{-17mm}
\caption{
Static quark potential obtained with \RC\ action with $c_{SW}={\rm pMF}$ on 
an $16^3{\times}32$ lattice.}  
\label{fig:VvsR16x32}
\vspace{-6mm}
\end{figure}

Another interesting question one can address with data for the static 
potential is whether the lattice scale 
determined from the string tension $\sigma$ is consistent with that from 
the $\rho$ meson mass.  

We extract the string tension assuming the form
$V(r)=V_0-{\alpha}/r+{\sigma}{\cdot}r$.  
The Coulomb coefficient $\alpha$ is ill-determined with our data, 
which starts at a coarse distance of $r_{\mbox{\scriptsize min}}=a\approx 0.2$~fm. 
We therefore fix $\alpha=0.3$, where ${\chi}^2/{\rm dof}$ is generally 
the smallest, and use the shift of $\sigma$ over the range $\alpha=0.2-0.4$ 
as a systematic error.  This fitting is repeated over several fitting ranges, 
and the average of the fitted results is taken as the central value of 
$\sigma$, while the variance is included into the systematic error. 
For estimating the lattice spacing 
we employ the phenomenological value ${\sigma}=(440{\rm MeV})^2$.
In Fig.~\ref{fig:a} we plot the lattice spacing obtained from
the string tension and the $\rho$ meson mass for each value of $K$ for all
four action combinations.  For the $\rho$ meson mass the value in the chiral 
limit is also shown.  

Let us emphasize that the consistency of the two 
determinations of the lattice scale is expected only in the chiral limit 
and sufficiently close to the continuum limit. 
From this point of view, a significant discrepancy in the chiral limit between 
the two determinations, as observed in the \PW\ and \RW\ results,
shows the presence of large scaling violating effects,
originating from the Wilson quark action, at $a^{-1}\approx 1$~GeV. 
We expect the discrepancy to disappear closer to the continuum
limit, and this is supported by the results obtained 
at $\beta=5.5$ with $a^{-1}\approx 2$~GeV of Ref.~\cite{SCRI-PW}.  

In contrast, the two estimates of the lattice spacing converge well 
to a consistent result for the \PC\ and \RC\ cases.
This shows that the clover term helps to improve the consistency of the
two determinations of the lattice spacing already at $a^{-1}\approx
1.0$GeV.

\subsection{Sea quark effects in the potential}

Our results for the potential, plotted in Fig.~\ref{fig:Vcrv}, and covering 
distances up to $r\approx 1$~fm, are quite linear in 
$r$ and do not show any sign of string breaking via $q\overline{q}$ 
pair creation.
In order to search for this effect, 
we employ one of the \RC\ runs carried out on a $16^3\times 32$ lattice
at $\beta=1.9$ with which we can measure the potential up to 
$r \simeq 2 \mbox{fm}$.  The results are shown in Fig.~\ref{fig:VvsR16x32},
and still do not show any sign of deviation from a linear behaviour in $r$.  

We find, however, that the maximum overlap of the 
ground state with the smeared Wilson loop operator 
decreases as $r$ increases 
and falls down to about 50\% at $r \simeq 2.0$~fm for our full QCD 
case, while effectively a 100\% overlap is obtained for the quenched case,
independent of the distance $r$. 
This may be an indication that the ground state develops towards larger $r$ a mixing,
absent in quenched QCD, with states containing $q\overline{q}$ 
pairs.

\begin{figure}[t]
\vspace{-3mm}
\begin{center}
\leavevmode
\epsfxsize=7.5cm
\epsfbox{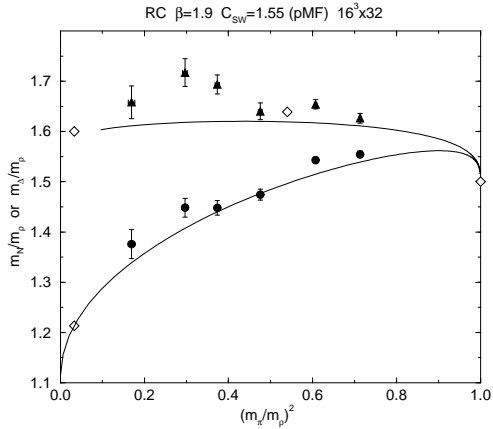}
\end{center}
\vspace{-17mm}
\caption{Results on a 
$16^3\!\times\!32$ lattice for \RC\ action at $\beta=1.9$.
Diamonds are experimental points corresponding to ${\rm
N}(940)/\rho(770)$, $\Delta(1232)/\rho(770)$ and $\Omega(1672)/\phi(1020)$.
}
\label{fig:16x32}
\vspace{-6mm}
\end{figure}

\section{Towards the chiral limit}

A very interesting and important issue in simulations in full QCD 
is how close one can approach the chiral limit.  
Of the four action combinations we have examined, the \RC\ combination 
exhibits the most improved behaviour with regard to the scaling of 
mass ratios and the rotational symmetry of the potential.   
We therefore choose this action combination for a feasibility 
study towards the chiral limit.
The runs are made at $\beta=1.9$ on a $16^3\!\times\!32$ lattice 
with a spatial size of about $2.6$~fm, and a quark mass corresponding 
to $m_\pi/m_\rho$ as small as $\approx 0.4$ is explored (see
Table~\ref{tab:param}).   
Results for the mass ratios obtained with 80--260 configurations 
for each quark mass, which includes statistics accumulated after 
the conference (especially for the lightest quark), are shown in 
Fig.~\ref{fig:16x32}. 


These results point towards the possiblity that, 
by improving both the fermion and the gauge
action, continuum results in full QCD at small quark masses are achievable
using the presently available computing power.

\vspace*{9mm}

This work is supported in part by the Grants-in-Aid
of Ministry of Education, Science and Culture
(Nos.\ 08NP0101, 08640349, 08640350, 08640404, 08740189, 
and 08740221).  Three of us (GB, RB, and TK) are supported by 
the Japan Society for the Promotion of Science.

\end{document}